# Hierarchical Re-ranker Retriever (HRR)
A Multi-level Chunking and Context-aware Reranking Framework for Efficient Document Retrieval

Ashish Singh, Priti Mohapatra
watsonx Client Engineering, IBM

## 1. Abstract

Retrieving the right level of context for a given query is a perennial challenge in information retrieval—too large a chunk dilutes semantic specificity, while chunks that are too small lack broader context. This paper introduces the **Hierarchical Re-ranker Retriever (HRR)**, a framework designed to achieve both fine-grained and high-level context retrieval for large language model (LLM) applications. In HRR, documents are split into **sentence-level** and **intermediate-level** (512 tokens) chunks to maximize vector-search quality for both short and broad queries. We then employ a reranker that operates on these 512-token chunks, ensuring an optimal balance—neither too coarse nor too fine—for robust relevance scoring. Finally, top-ranked intermediate chunks are mapped to **parent chunks** (2048 tokens) to provide an LLM with sufficiently large context.

We compare HRR against three widely used alternatives(details of them can be found in appendix section):

1. **Base Retriever + Reranker**
2. **ChildToParent(C2P) Retriever + Reranker**
3. **SentenceToParent(S2P) Retriever + Reranker**

Experiments on two datasets—**Yojana** and **Lendryl**—demonstrate that **HRR** consistently outperforms these baselines in both **Hit Rate (HR)** and **Mean Reciprocal Rank (MRR)**. On Yojana, HRR achieves a perfect **100% Hit Rate** and an MRR of **96.15%** which is **25% higher** than Base Retriever and around **15% higher** than C2P or S2P retriever. Similarly, on Lendryl, HRR attains MRR which is **20% and 10% higher** than Base Retriever and C2P or S2P retriever respectively.

These results confirm that a multi-stage retrieval strategy—fine-grained sentence-level and intermediate level(512 token) filtering, optimized 512 token reranking, and final parent-chunk(2048 token) mapping—delivers more accurate, context-rich retrieval well-suited for downstream LLM tasks.

## 2. Acknowledgement

We thank IBM for generously supporting this re-search by providing the necessary computing infrastructure, and LLM resources.

## 3. Introduction

In recent years, large language models (LLMs) have significantly improved the quality of natural language understanding and question answering tasks. However, delivering highly relevant context to these models remains a non-trivial challenge, especially when dealing with lengthy documents. **Chunk-based retrieval**—splitting long texts into smaller units and indexing them—has emerged as a common solution. Despite its success, the question of choosing an **optimal chunk size** for both retrieval and reranking continues to be a subject of debate.

On one end of the spectrum, **large chunks** (e.g., 2048 tokens) capture broader context and can serve LLMs better, allowing them to see the "big picture." Yet such large chunks often produce **generic embeddings**, as the sheer volume of text dilutes domain-specific keywords and relevant subtopics. On the other end, **very small chunks**, such as sentences, enable precise matching of short or domain-specific queries but lose the valuable context that a paragraph or section might provide.

A second challenge arises in **reranking**. Neural rerankers typically work best on "moderate-sized" text segments—chunks that are neither too large to overshadow local relevance signals, nor too small to omit key context. Empirical evidence suggests that **512-token** chunks are well-suited for reranker models, effectively balancing context-richness and retrievability. Existing retrieval approaches often overlook this balance, leading to one of the following drawbacks:

1. **Base Retriever + Reranker**: Uses a single, large chunk size (2048 tokens) for both retrieval and reranking, yielding coarse, generic embeddings and suboptimal reranking quality.
2. **C2P Retriever + Reranker**: Traverses a hierarchy (e.g., 2048 → 512 → 256), but reranking occurs at the 2048 level. While it partially addresses granularity issues, the final reranking does not capitalize on the more context-friendly 512-token scale.
3. **S2P Retriever + Reranker**: Focuses entirely on sentence-level retrieval but again handles reranking at 2048 tokens. Although it can capture domain-specific keywords, it fails to exploit an intermediate chunk size that might better reflect broader context for ranking relevance.

To address these limitations, we propose the **Hierarchical Re-ranker Retriever (HRR)**, which **simultaneously** exploits sentence-level and intermediate-level (512 tokens) chunks for retrieval, then reranks at the **512-token level**, and finally returns top-ranked parent chunks (2048 tokens) for enhanced contextual coverage. Specifically, the HRR workflow is:

1. **Initial Retrieval**: Retrieve the most similar sentence-level and intermediate-level chunks.
2. **Reranking**: Use a neural reranker model to score intermediate chunks (512 tokens), striking an ideal size compromise for relevance estimation.
3. **Parent Chunk Linking**: Map the top-ranked intermediate chunks back to their parent chunks (2048 tokens), ensuring the LLM has sufficient context to produce comprehensive answers.

By combining the fine-grained precision of sentence-level matches with an optimally sized reranking process and ultimately providing large, context-rich parent chunks to the LLM, HRR addresses a gap in existing methods. As we demonstrate in later sections, this approach achieves strong performance gains—most notably in Mean Reciprocal Rank (MRR) and Hit Rate (HR)—across diverse datasets such as Yojana and Lendryl, underscoring HRR's effectiveness for both short keyword-based and more expansive queries.

## 4. Background & Related Work

The field of information retrieval (IR) has undergone remarkable advancements in the past decade, driven by innovations in neural embeddings, approximate nearest neighbour (ANN) search, and large language models (LLMs). This section reviews existing literature on chunk-based retrieval, hierarchical parsing strategies, and neural re-ranking, illustrating how our Hierarchical Re-ranker Retriever (HRR) extends or differs from these approaches.

### 4.1 Chunk-Based Retrieval

**Document vs. Chunk-Level Retrieval**
Traditional IR systems, such as BM25 or TF–IDF, typically operate on an entire document representation. While computationally efficient, this approach can be overly coarse for large documents covering multiple topics. As a result, **chunk-based retrieval** emerged to handle long texts more effectively by splitting them into manageable segments—paragraphs, sentences, or fixed-size token windows. Dense Passage Retrieval (DPR) and other neural methods have shown that embedding these smaller chunks often yields more fine-grained matching, ultimately improving retrieval quality for focused queries.

**Optimal Chunk Size**
Determining the ideal chunk length is still a subject of debate. Short segments (e.g., individual sentences) capture domain-specific keywords and align well with short or domain-focused queries. However, they may lose broader context. Conversely, large chunks (e.g., ~1000 tokens or more) provide greater context but risk producing generic embeddings that dilute salient keywords. Many systems solve this problem by choosing a single chunk size that best fits their domain, often ignoring the potential benefit of combining multiple chunk sizes.

## 4.2 Hierarchical Retrieval Approaches

**Layered or Recursive Chunking**
Various hierarchical retrieval methods have been proposed to exploit multiple levels of granularity within a document. A document might be divided into sections (parent level), paragraphs (child level), and sentences (leaf level). Approaches such as *C2P Retriever + Reranker* attempt to retrieve large chunks first and then drill down into sub-chunks. Although these methods capture different resolutions of context, they often lack an integrated reranking mechanism that effectively leverages mid-sized chunks (e.g., 512 tokens) for relevance scoring.

**Context Preservation**
Hierarchical chunking is particularly useful in tasks like question answering (QA), where local detail must be understood within a broader section. Nevertheless, the transition between different chunk sizes can introduce discontinuities if not carefully designed. Some methods mitigate this by overlapping segments or by maintaining detailed metadata that links child chunks back to their parents. Despite these improvements, many hierarchical systems ultimately rerank using the largest chunks, missing the advantages of a more moderate chunk size for relevance estimation.

## 4.3 Neural Re-Ranking

**Initial Retrieval vs. Reranking**
A standard pipeline in modern IR consists of two stages: (1) **initial retrieval** with a lightweight similarity measure (like dense embeddings or BM25) to shortlist candidates, and (2) **reranking** with a computationally heavier model (e.g., a transformer-based cross-encoder). This two-stage paradigm leverages the speed of ANN search for candidate generation and the accuracy of deep networks for final ranking.

**Chunk Size in Re-Ranking**
Although recent neural rerankers (e.g., BERT-based or T5-based models) can handle up to ~512 tokens effectively, larger inputs (e.g., 2048 tokens) may lead to truncated context or reduced scoring precision if the model is not specifically optimized for long-form inputs. In many existing systems, the chunk size used in reranking is inherited from the chunk size used in retrieval or from the maximum sequence length allowed by the model—both of which can be suboptimal.

## 4.4 Positioning Our Work

Most current methods adopt **one** of the following strategies:

1. **Single-Chunk Baseline (2048)**: Embedding and reranking at 2048 tokens yields broad context but compromises on embedding specificity.
2. **Hierarchical Parsers (2048 → 512 → 256)**: While they parse the document at multiple levels, they often finalize ranking decisions at the largest chunk size (2048), limiting fine-grained relevance.
3. **Sentence-Level Retrieval**: Achieves high keyword specificity but may lose essential contextual clues, especially if the reranker also operates at an overly large or overly small scale.

Our **Hierarchical Re-ranker Retriever (HRR)** addresses these gaps by:

- **Combining Multiple Chunk Levels**: Sentences (fine-grained) & intermediate chunks (512 tokens) for initial retrieval and intermediate chunks (512 tokens) for reranking.
- **Maximizing Context**: Mapping the top-ranked intermediate chunks back to larger parent chunks (2048 tokens) ensures that the final answer context is sufficiently broad for LLM input.
- **Balancing Efficiency and Relevance**: Using sentence-level vectors for candidate generation remains efficient, while re-ranking at the 512-token level aligns well with transformer-based scoring mechanisms.

This approach not only preserves local lexical cues but also maintains sufficient contextual information, leading to superior Hit Rate (HR) and Mean Reciprocal Rank (MRR). The next section elaborates on the design and implementation of HRR, detailing the multi-level retrieval, chunk embedding, and reranking processes that form the core of our system.

## 5. Methodology

In this section, we present the **Hierarchical Reranker Retriever (HRR)**, which processes documents into a hierarchical structure of chunks—parent, intermediate, and sentence-level—to balance retrieval efficiency and contextual relevance. Figure 1 provides a high-level overview of the architecture.

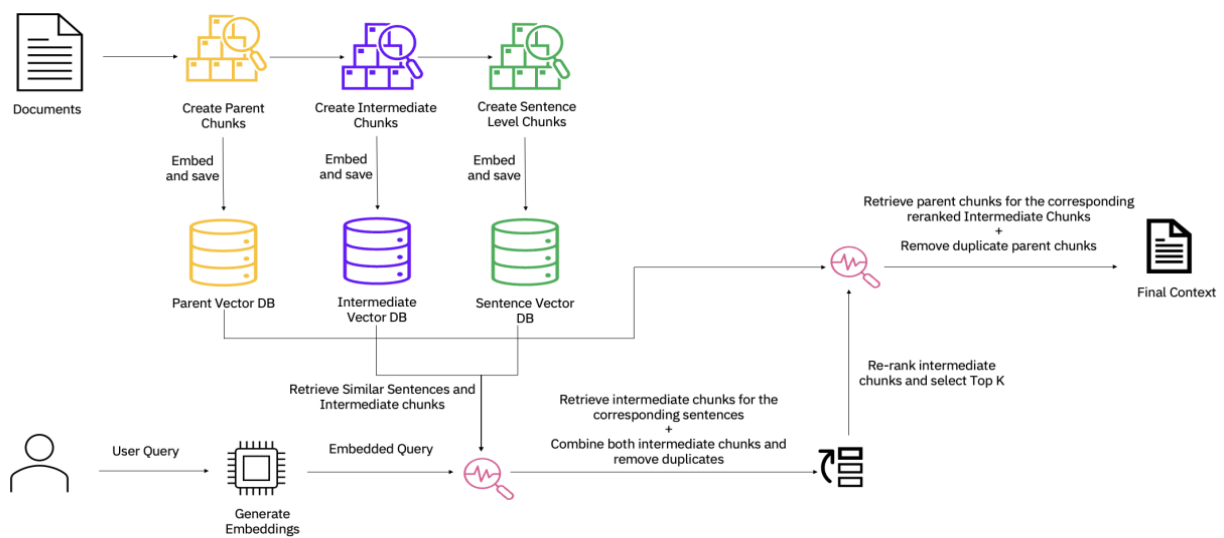

Figure 1 – HRR Architecture Diagram

## 5.1 Overview

The HRR method is designed to maximize retrieval relevance by capturing and preserving context at multiple levels of granularity. The framework comprises the following key steps:

1. **Hierarchical Chunking:** Each document is split into three granularities—parent chunks (chunk size 2048), intermediate chunks (chunk size 512), and sentence-level chunks.
2. **Embedding and Indexing:** All chunks (at each level) are embedded using a pretrained language model and stored in a vector database.
3. **Multi-level Retrieval**: Top-k sentence-level and intermediate-level chunks are retrieved for a given query.
4. **Reranking**: A neural re-ranker model re-scores the retrieved intermediate chunks to highlight the most relevant ones.
5. **Parent Chunk Mapping**: The highest-ranked intermediate chunks are mapped back to their parent chunks.
6. **Final Retrieval**: A set of unique parent chunks is returned as the final, context-rich answer.

This hierarchical approach ensures that the granularity of sentence-level retrieval is combined with broader contextual insight from intermediate and parent chunks.

## 5.2 Hierarchical Chunking

To preserve both local detail and global coherence, each document is divided into three levels of chunks:

1. **Parent-Level Chunking**
    - Each document is initially split into **parent chunks** of up to 2048 tokens. This size is chosen to capture broader context, such as entire sections or chapters, while keeping the chunks within a manageable token limit.
2. **Intermediate-Level Chunking**
    - Each parent chunk is further segmented into **intermediate chunks** of up to 512 tokens. Intermediate chunks serve as a balance between large, context-rich segments and finer-grained text passages.
3. **Sentence-Level Chunking**
    - Finally, each intermediate chunk is split into **sentence-level chunks**. Sentence boundary detection ensures even more granular representation for embedding.

Throughout the chunking process, metadata links each sentence to its intermediate chunk and each intermediate chunk to its parent chunk, forming a complete hierarchy.

## 5.3 Embedding and Indexing

After chunking, **all** chunks are embedded using the pretrained language model **BAAI/bge-base-en-v1.5**, which is optimized for semantic understanding. Formally, for a text chunk t, we obtain an embedding vector e by:

$$e = \mathrm{Embed}(t),$$

where Embed(·) is the encoder's forward pass.

All parent, intermediate, and sentence-level chunks are indexed in a vector database (e.g., FAISS, Milvus, or Elasticsearch with vector search). Each chunk is stored alongside its embedding vector and metadata indicating its position in the hierarchy. Indexing at multiple levels allows for flexible retrieval strategies that can query the database at different granularities.

## 5.4 Query Retrieval Process

Given a query Q, the HRR framework performs a multi-level retrieval to capture both granular and broader context:

1. **Embedding the Query**

    - The query is converted into an embedding q using the same pretrained model for consistency.

2. **Multi-level Retrieval**

    - Using q, the system retrieves the **top-k sentence-level chunks** and **top-k intermediate-level chunks** from the vector database, based on cosine similarity scores.
    - This ensures coverage of both fine-grained matches (sentence chunks) and medium-grained matches (intermediate chunks).

3. **Mapping to Intermediate Chunks**

    - Each retrieved sentence-level chunk is mapped to its corresponding intermediate chunk, preserving the hierarchical relationships established during the chunking process.

4. **Duplicate Removal**

    - To avoid redundancies, duplicate intermediate chunks arising from multiple retrieved sentences are removed, leaving a set of unique intermediate chunks linked to the query.

## 5.5 Reranking

Next, the retrieved intermediate chunks are reranked to highlight those most relevant to Q. We construct an input pair consisting of:

$$(\mathbf{Q}, \mathbf{C}_{int})$$

where **C**int is the embedding or text representation of an intermediate chunk.

### 5.5.1 Scoring Function

We use a pretrained reranker model, **jinaai/jina-reranker-v1-turbo-en**, which takes a query–chunk pair (**Q**, **C**int) and outputs a relevance score:

$$s(\mathbf{Q}, \mathbf{C}_{int}) = \text{Reranker}(\mathbf{Q}, \mathbf{C}_{int}).$$

where higher scores indicate stronger relevance. The re-ranker can leverage additional contextual signals, such as aggregated similarity from its contained sentence chunks.

*5.5.2 Top-k Selection*

We sort the intermediate chunks based on their relevance scores and select the top-k intermediate chunks:

$$\{c_{int}^1, c_{int}^2, \ldots, c_{int}^k\}$$

Each chosen intermediate chunk is expected to contain or relate to highly relevant information.

## 5.6 Parent Chunk Mapping

The **top-K** intermediate chunks after reranking are mapped to their respective **parent chunks**. As with intermediate chunks, duplicates are removed to maximize diversity and reduce redundancy in the final output. This step reintroduces the wider context in which each highly relevant intermediate chunk is situated.

## 5.7 Final Retrieval

The resulting **unique parent chunks** derived from the highest-ranked intermediate chunks form the final retrieval set. By returning parent chunks, users receive comprehensive context along with the key information that matched their query. This hierarchical approach—starting with fine-grained sentence-level retrieval and culminating in a smaller set of context-rich parent chunks—achieves a balance between precision and completeness.

**In summary**, the HRR framework leverages a three-tier chunking strategy, a specialized sentence-embedding model for fine-grained retrieval, and a reranker model to refine intermediate chunks. The final parent chunks ensure that users receive results that are both accurately targeted and contextually rich. In the next section, we detail the experimental setup and metrics used to validate the effectiveness of this framework.

# 6. Experimental Setup

This section describes the datasets, chunking parameters, re-ranker configuration, and evaluation metrics used to assess the Hierarchical Reranker Retriever (HRR) framework. We performed experiments on **Yojana** and **Lendryl** datasets to demonstrate the effectiveness of HRR in retrieving contextually relevant information from large documents.

## 6.1 Datasets

1. **Yojana**: A collection of government policy articles, typically spanning multiple sections or sub-sections.
2. **Lendryl**: A repository of domain-specific technical literature characterized by diverse sections and subtopics.

Both datasets contain long documents well-suited for hierarchical chunking, allowing us to evaluate how effectively HRR preserves both fine-grained and global context.

## 6.2 Chunking Parameters

To enable hierarchical retrieval, documents were chunked at three levels: **base (parent)**, **intermediate**, and **sentence**. Table I (below) summarizes the chunking parameters.

| Parameters | Value |
|---|---|
| Embed model | BAAI/bge-small-en |
| Base chunk size | 2048 |
| Base chunk overlap | 0 |
| Intermediate chunk size | 512 |
| Intermediate chunk overlap | 0 |
| Similarity top k | 10 |
| Rerank top k | 5 |

**Table I: Chunking Parameters for Experimentation**

1. **Base (Parent) Chunk**: Each document is split into 2048-token segments with an overlap of 0 tokens. Ideally we should use chunk overlap of 200 but for this experiment we used 0 so that multiple chunks does not have same answer.
2. **Intermediate Chunk**: Each base chunk is further divided into 512-token segments (no overlap). These medium-grained chunks help preserve local context while reducing the length of text segments.
3. **Sentence-Level Chunk**: Each intermediate chunk is split into individual sentences to achieve fine-grained embeddings and more precise retrieval.

## 6.3 Embedding and Retrieval Models

- **Embedding Model**: We used a locally hosted instance of BAAI/bge-small-en to generate vector embeddings for sentence-level chunks. This model is optimized for semantic understanding and provides efficient encoding of short text segments.
- **Reranker Model**: The jinaai/jina-reranker-v1-turbo-en model from Hugging Face was employed to re-score intermediate chunks after initial retrieval. This reranker assigns a relevance score to each intermediate chunk based on the query–chunk relationship.

## 6.4 Context and Question Pair Generation

For part of our evaluation, we generated synthetic query–context pairs using **mistralai/mistral-large**. The model configuration is shown in Table II:

| Model | Parameters |
|---|---|
| mistralai/mistral-large | Decoding Method: Greedy<br>Max New Tokens: 1500<br>Min New Tokens: 1 |

**Table II: Model Parameters for Context and Question Pair Generation**

## 6.5 Evaluation Metrics

Conventional retrieval metrics do not fully capture the hierarchical and contextual aspects of HRR. Consequently, we employed two tailored metrics to measure both accuracy and ranking quality:

1. **Hit Rate (HR)**

$$\text{HR} = \frac{\text{Number of successful hits in top-K results}}{\text{Total Queries}}$$

   The Hit Rate (HR) measures how frequently the correct parent chunk appears among the top-K retrieved results. It evaluates the system's ability to prioritize relevant information effectively.

2. **Mean Reciprocal Rank (MRR)**

$$\text{MRR} = \frac{1}{N} \sum_{i=1}^{N} \frac{1}{\text{Rank of the first relevant document or chunk for query } i}$$

   MRR focuses on the position of the first relevant chunk. A higher MRR indicates that relevant chunks consistently appear near the top of the ranked list.

## 6.6 Compared Retrieval Mechanisms

We compared **HRR** against three popular retrieval setups on both the **Yojana** and **Lendryl** datasets:

1. **Base Retriever + Reranker**: A single-chunk baseline with default settings, relying on large (2048-token) segments for both retrieval and reranking.

2. **C2P Retriever + Reranker**: A top-down hierarchical parser (e.g., 2048→512→256) designed to capture multi-level context but still reranks at the largest chunk size.

3. **S2P Retriever + Reranker**: A fine-grained parser that breaks text into sentence-level segments for higher precision but lacks an intermediate layer for reranking.

4. **Results_Chunk_HRR (Proposed)**: Our hybrid method combining sentence-level and intermediate retrieval, a neural reranker at intermediate(512-token) chunks, and final mapping to 2048-token parent chunks for comprehensive context.

## 6.7 Experiment Procedure

1. **Document Preparation**: Long documents from both datasets were chunked into three layers (base, intermediate, and sentence) using the parameters in Table I.
2. **Indexing**: Sentence-level embeddings were indexed using a vector store (e.g., llama_index) for efficient similarity-based retrieval.
3. **Initial Retrieval**: Each query was embedded and matched against the vector store to retrieve the top-10 most similar sentences and intermediate chunks.
4. **Reranking**: We utilized the Jina reranker to reorder the intermediate chunks based on relevance scores.

5. **Final Output**: The top-5 reranked intermediate chunks were mapped back to their parent chunks to provide comprehensive context in the final retrieval.

In summary, this experimental setup ensures that HRR is evaluated comprehensively on real-world data and against multiple baselines. The next section (Section 7) provides the results obtained from these experiments, along with discussions and insights into the strengths and limitations of the proposed framework.

# 7. Results

In this section, we report the effectiveness of our **Hierarchical Reranker Retriever (HRR)** on the **Yojana** and **Lendryl** datasets, comparing it against three alternative retrieval methods: the Base Retriever + Reranker, C2P Retriever + Reranker, and S2P Retriever + Reranker. We evaluate each approach using **Hit Rate (HR)** and **Mean Reciprocal Rank (MRR)** to quantify both coverage (whether relevant chunks appear in the top-K results) and ranking quality.

## 7.1 Quantitative Results

**Table III** shows the performance on the Yojana dataset, while **Table IV** summarizes results on Lendryl.

| Retriever | Hit Rate | MRR |
|---|---|---|
| **Base Retriever + Reranker** | 0.897436 | 0.717521 |
| **C2P Retriever + Reranker** | 1.000000 | 0.819231 |
| **S2P Retriever + Reranker** | 0.974359 | 0.784188 |
| **Results_Chunk_HRR (Proposed)** | **1.000000** | **0.961538** |

Table III: Comparison on Yojana Dataset

| Retriever | Hit Rate | MRR |
|---|---|---|
| **Base Retriever + Reranker** | 0.736111 | 0.636343 |
| **C2P Retriever + Reranker** | 0.930556 | 0.748611 |
| **S2P Retriever + Reranker** | 0.958333 | 0.751157 |
| **Results_Chunk_HRR (Proposed)** | **0.958333** | **0.847220** |

Table IV: Comparison on Lendryl Dataset

- **Yojana**: Both ParentToChildrenNodeParser and HRR achieve a perfect **Hit Rate (1.0)**, but HRR substantially outperforms in **MRR (0.9615 vs. 0.8192)**, indicating that relevant chunks rank higher.
- **Lendryl**: HRR posts the highest **MRR (0.8472)** among all methods, confirming that it not only retrieves the correct chunks but also consistently ranks them closer to the top.

## 7.2 Discussion of Key Findings

A. **Effectiveness of Hierarchical Chunking**
   Splitting each document into **parent (2048 tokens), intermediate (512 tokens), and sentence-level** chunks preserves both local detail and global context. Single-granularity approaches (e.g., Base Retriever) either lose fine-grained keyword matching or dilute context in overly large chunks.

B. **Role of the Reranker**
Incorporating a neural reranker on **512-token** chunks significantly boosts MRR. Approaches that rerank solely at the 2048-token level (Base Retriever + Reranker, C2P Retriever + Reranker, S2P Retriever + Reranker) miss the optimal mid-grained context for precise scoring.
C. **Balanced Retrieval Strategy**
   a. S2P Retriever + Reranker offers fine-grained sentence retrieval but reranks at 2048 tokens, limiting its MRR improvements.
   b. **HRR** strikes a better balance by capturing short-range relevance signals at the sentence level & mid-range relevance signals at the intermediate level, reranking them in the 512-token intermediate context, then returning the parent (2048 tokens) for comprehensive answers.

## 7.3 Qualitative Observations

- **Base Retriever + Reranker** often returns large blocks of text containing partial or tangentially relevant information.
- **C2P Retriever + Reranker** captures more granularity but still suffers from final reranking on large parent chunks, reducing precision.
- **S2P Retriever + Reranker** excels at pinpointing precise sentences but can miss intermediate context crucial for accurate ranking.
- **HRR** seamlessly blends these advantages by using **sentence-level** and **intermediate level(512)** matching for high precision, **512-token** reranking for context-rich relevance, and **2048-token** parent chunks for comprehensive coverage.

## 7.4 Summary of Advantages

1. **High Precision Early Ranking**: Fine-grained sentence and intermediate chunk retrieval combined with a 512-token reranker ensures the most relevant chunks are placed at the top, reflected in strong MRR gains.
2. **Context Preservation**: Providing 2048-token parent chunks to the end-user or LLM retains broader topical context, reducing disjointed answers.
3. **Scalability**: The hierarchical design narrows down candidate chunks at each stage, making it feasible even for large documents.

In conclusion, **HRR** demonstrates superior performance over existing baselines, offering an optimal balance between fine-grained relevance and broader context—a critical requirement for downstream LLM tasks such as question answering and in-depth document exploration.

## 8. Limitations and Future Work

Despite its strong performance, the **Hierarchical Reranker Retriever (HRR)** framework has several areas where further research and optimization could be beneficial:

1. **Storage and Computational Overhead**

    - **Multi-Level Embeddings**: Storing embeddings at the sentence, intermediate and parent levels increases indexing complexity and storage requirements. Future work could

explore more compact embedding strategies (e.g., dimensionality reduction or sparse representations) or using one vector dB to mitigate this overhead.

2. **Static Chunk Boundaries**

    - **Fixed Token Sizes**: Relying on fixed chunk sizes (2048, 512, and sentence-level) may not be optimal for all document types or domain requirements. Adaptive chunking strategies—driven by content structure or query type—could further refine retrieval accuracy.

3. **Domain Adaptation**

    - **Limited Dataset Variety**: While HRR was evaluated on the Yojana (policy) and Lendryl (technical) datasets, its generality across different domains (e.g., legal, biomedical) and languages is not fully tested. Incorporating domain adaptation or multilingual models could widen the applicability of HRR.

4. **Handling Overlapping Context**

    - **Minimal Overlap Approach**: Current settings favor minimal overlap (e.g., 200 tokens at the base level, 0 at intermediate) to reduce redundancy. However, some queries may benefit from higher overlap for improved continuity. A dynamic overlap strategy may address this trade-off.

## Future Work

1. **Adaptive Chunking**: Investigate chunk boundary decisions made on-the-fly based on document structure, query length, or semantic content.
2. **Cross-Lingual Retrieval**: Extend HRR to support multilingual corpora and cross-lingual queries, a growing need in global IR systems.
3. **On-Device Inference**: Optimize HRR components (e.g., through model distillation or quantization, using single vector DB instead of 3 instances) for deployment in resource-constrained environments.

By addressing these challenges, future iterations of HRR can further improve retrieval quality, scalability, and applicability across diverse use cases.

## 9. Conclusion

In this paper, we introduced the **Hierarchical Reranker Retriever (HRR)** to address the long-standing challenge of balancing fine-grained retrieval precision with broader contextual coverage. By segmenting documents into three levels—sentence, intermediate (512 tokens), and parent (2048 tokens)—HRR preserves key domain-specific signals while ensuring a sufficiently large context for downstream LLM tasks. The reranking step at the 512-token level further refines relevance, overcoming the limitations of both overly large chunks (leading to generic embeddings) and tiny chunks (losing context).

Extensive experiments on the **Yojana** and **Lendryl** datasets show that HRR substantially outperforms existing methods in Hit Rate and Mean Reciprocal Rank. These gains underscore the importance of a middle-layer chunk size for reranking, along with precise sentence-level candidate retrieval and final parent-chunk aggregation. Despite the additional storage and indexing overhead, HRR's hybrid approach consistently demonstrates high retrieval accuracy and robust context preservation—indicating that a careful hierarchical strategy is crucial for modern large language model pipelines.

## 10. References

- R. Nogueira and W. Lin, "From rank to re-rank: Learning a neural ranker for dense retrieval," in *Proc. 2020 Conf. Empirical Methods in Natural Lang. Processing (EMNLP)*, Online, 2020, pp. 8137–8146. doi: 10.18653/v1/2020.emnlp-main.654.
- X. Liu, J. Ma, and J. Gao, "Hierarchical dense retrieval with multi-level chunking for long documents," in *Proc. 2023 Conf. Neural Information Processing Systems (NeurIPS)*, New Orleans, LA, USA, 2023.
- "Retrieve & Re-Rank — Sentence Transformers," *SBERT Documentation*, 2023. [Online]. Available: https://sbert.net/examples/applications/retrieve_rerank/
- J. Glorat, "Context-aware chunking for enhanced retrieval-augmented generation," *Medium*, Oct. 23, 2023. [Online]. Available: https://medium.com/%40glorat/context-aware-chunking-for-enhanced-retrieval-augmented-generation-oct23-9dcd435d9cf1
- "Chunking strategies for LLM applications," *Pinecone*, 2024. [Online]. Available: https://www.pinecone.io/learn/chunking-strategies/
- Y. Liu, H. Wu, and S. Wang, "Cooperative training of retriever-reranker for effective dialogue response selection," in *Proc. 61st Annual Meeting of the Association for Computational Linguistics (ACL)*, 2023, pp. 3101–3112. doi: 10.18653/v1/2023.acl-long.174.
- S. Thakur, N. Reimers, and I. Gurevych, "Large language models for information retrieval: A survey," *arXiv preprint arXiv:2308.07107*, 2023. [Online]. Available: https://arxiv.org/abs/2308.07107.
- H. Zhu, "Reranking passages with coarse-to-fine neural retriever enhanced by list-context information," *arXiv preprint arXiv:2308.12022*, 2023. [Online]. Available: https://arxiv.org/abs/2308.12022.
- Y. Zhang, D. Long, G. Xu, and P. Xie, "HLATR: Enhance multi-stage text retrieval with hybrid list aware transformer reranking," *arXiv preprint arXiv:2205.10569*, 2022. [Online]. Available: https://arxiv.org/abs/2205.10569.
- N. Gu, Y. Gao, and R. H. R. Hahnloser, "Local citation recommendation with hierarchical-attention text encoder and SciBERT-based reranking," *arXiv preprint arXiv:2112.01206*, 2021. [Online]. Available: https://arxiv.org/abs/2112.01206.
- S. Xu, L. Pang, J. Xu, H. Shen, and X. Cheng, "List-aware reranking-truncation joint model for search and retrieval-augmented generation," *arXiv preprint arXiv:2402.02764*, 2024. [Online]. Available: https://arxiv.org/abs/2402.02764.
- X. Liu, J. Ma, and J. Gao, "Hierarchical dense retrieval with multi-level chunking for long documents," in *Proc. 2023 Conf. Neural Information Processing Systems (NeurIPS)*, 2023.
- Y. Zhang, H. Zhao, M. Chen, and J. Wang, "Hybrid hierarchical retrieval for open-domain question answering," in *Proc. 61st Annual Meeting of the Association for Computational Linguistics (ACL)*, 2023.
- R. Nogueira and W. Lin, "From rank to re-rank: Learning a neural ranker for dense retrieval," in *Proc. 2020 Conf. Empirical Methods in Natural Lang. Processing (EMNLP)*, 2020, pp. 8137–8146. doi: 10.18653/v1/2020.emnlp-main.654.
- Y. Sun, H. Deng, and X. Ren, "Dense hierarchical retrieval for open-domain question answering," in *Findings of the Association for Computational Linguistics (EMNLP)*, 2021. [Online]. Available: https://aclanthology.org/2021.findings-emnlp.19.pdf.

# 11. Appendices

**Base Retriever + Reranker** : This approach used base node parser which creates chunks of chunk size 2048. For a given query, these large chunks are retrieved based on similarity and then passed to a reranker for scoring. The reranked chunks are subsequently presented to the LLM for final answer generation.

**ChildToParent(C2P) Retriever + Reranker** : In this we used HierarchicalNodeParser from llama index which outputs a hierarchy of nodes(2048→512→256), from top-level nodes with bigger chunk sizes to child nodes with smaller chunk sizes, where each child node has a parent node with a bigger chunk size. Then for a given query we match all the nodes based on similarity and for any child node matched we get its parent node so the final output is the biggest parent node(2048) and sends those parent nodes to a reranker. The highest-scoring parents are then passed to the LLM to produce answers.
HierarchicalNodeParser -
https://docs.llamaindex.ai/en/stable/api_reference/node_parsers/hierarchical/

**SentenceToParent(S2P) Retriever + Reranker** : Building on the HierarchicalNodeParser, we developed a parser that generates a hierarchy from 2048-token parent nodes down to sentence-level nodes. Each sentence node maps to a single parent node. For a given query, sentence nodes are matched first; their respective 2048-token parents are then retrieved and reranked. Finally, the top-ranked parent nodes are provided to the LLM for answer generation.